\documentclass[english,preprintnumbers,amsmath,amssymb,superscriptaddress]{revtex4}
\usepackage[T1]{fontenc}
\usepackage[latin1]{inputenc}
\usepackage{color}
\usepackage{babel}
\usepackage{amsmath}
\usepackage{amssymb}
\usepackage{esint}
\usepackage[unicode=true,pdfusetitle,
 bookmarks=true,bookmarksnumbered=false,bookmarksopen=false,
 breaklinks=false,pdfborder={0 0 1},backref=section,colorlinks=false]
 {hyperref}
\hypersetup{
 hypertex}

\makeatletter
\@ifundefined{textcolor}{}
{%
 \definecolor{BLACK}{gray}{0}
 \definecolor{WHITE}{gray}{1}
 \definecolor{RED}{rgb}{1,0,0}
 \definecolor{GREEN}{rgb}{0,1,0}
 \definecolor{BLUE}{rgb}{0,0,1}
 \definecolor{CYAN}{cmyk}{1,0,0,0}
 \definecolor{MAGENTA}{cmyk}{0,1,0,0}
 \definecolor{YELLOW}{cmyk}{0,0,1,0}
 }

\@ifundefined{definecolor}
 {\usepackage{color}}{}

\usepackage{dcolumn}\usepackage{bm}\@ifundefined{definecolor}{\usepackage{color}}{}
\usepackage{amsfonts}\usepackage{amsthm}\usepackage{mathrsfs}\usepackage{subfigure}\usepackage{ulem}

\makeatother

\begin{document}

\title{A simple derivation of the Lindblad equation}

\author{Carlos Alexandre Brasil}

\affiliation{Instituto de Física \textquotedbl{}Gleb Wataghin\textquotedbl{},
Universidade Estadual de Campinas, Campinas, Brasil}

\email{carlosbrasil.physics@gmail.com}

\selectlanguage{english}%

\author{Felipe Fernandes Fanchini}

\affiliation{Faculdade de Ciências de Bauru, Universidade Estadual Paulista \textquotedbl{}Julio
de Mesquita Filho\textquotedbl{}, Bauru, Brasil}

\author{Reginaldo de Jesus Napolitano}

\affiliation{Instituto de Física de São Carlos, Universidade de São Paulo, São
Carlos, Brasil}
\begin{abstract}
We present a derivation of the Lindblad equation - an important tool
for the treatment of non-unitary evolutions - that is accessible to
undergraduate students in physics or mathematics with a basic background
on quantum mechanics. We consider a specific case, corresponding to
a very simple situation, where a primary system interacts with a bath
of harmonic oscillators at zero temperature, with an interaction Hamiltonian
that resembles the Jaynes-Cummings format. We start with the Born-Markov
equation and, tracing out the bath degrees of freedom, we obtain an
equation in the Lindblad form. The specific situation is very instructive,
for it makes it easy to realize that the Lindblads represent the effect
on the main system caused by the interaction with the bath, and that
the Markov approximation is a fundamental condition for the emergence
of the Lindbladian operator. The formal derivation of the Lindblad
equation for a more general case requires the use of quantum dynamical
semi-groups and broader considerations regarding the environment and
temperature than we have considered in the particular case treated
here. 
\end{abstract}
\maketitle
\begin{description}
\item [{{Keywords:}}] Lindblad equation, open quantum systems 
\item [{{PACS:}}] 03.65.Ca Formalism, 03.65.Yz Decoherence; open systems;
quantum statistical methods 
\end{description}

\section{Introduction}

The Lindblad equation \cite{key-1} is the most general form for a
Markovian master equation, and it is very important for the treatment
of irreversible and non-unitary processes, from dissipation and decoherence
\cite{key-2} to the quantum measurement process \cite{key-3,key-4}.
For the latter, in recent applications \cite{key-4,key-12}, the Lindblad
equation was used in the introduction of time in the interaction between
the measured system and the measurement apparatus. Then, the measurement
process is no longer treated as instantaneous, but finite, with the
duration of that interaction changing the probabilities - diagonal
elements of the density operator - associated to the possible final
results. On the other hand, in quantum optics, the analysis of spontaneous
emission on a two-level system conducts to the Lindblad equation \cite{key-13}.
At last, in the case of quantum Brownian movement, it is possible
to transform the Caldeira-Leggett equation \cite{key-14} into Lindblad
with the addition of a term that becomes small in the high-temperature
limit \cite{key-2}. These are a couple of many applications of the
Lindblad equation, justifying its understanding by students in the
early levels.

Contrasting against its importance and wide range of applications,
its original deduction \cite{key-1} involves the formalism of quantum
dynamical semigroups \cite{key-5,key-6}, which is quite unfamiliar
to most of the students and researchers. Other more recent ways to
derive it involve the use of Itô stochastic calculus \cite{key-7,key-8}
or, in the specific case of quantum measurements, considerations about
the interaction between the system and the meter \cite{key-9}. Another
deduction, where the quantum dynamical semigroups are not explicitly
used can be found on Ref. \cite{key-2}. These methods, their assumptions,
their applications, and, more importantly, their physical meanings
appear very intimidating to beginning students.

To make the Lindblad equation more understandable, this article presents
its deduction in the specific case of two systems: $S$, the principal
system, and $B$, which can be the environment or the measurement
apparatus, at zero temperature, with an interaction between them that
resembles the one of the Jaynes-Cummings model\cite{key-2}. Initially
we derive the Born-Markov master equation \cite{key-2} and then we
trace out the degrees of freedom of system $B$. The Lindbladian emerges
naturally as a consequence of the Markov approximation. Each Lindblad
represents the effect on system $S$ caused by the $S-B$ interaction.

Clearly, the present approach does not prove the general validity
of the Lindblad equation. Our intention is simply to provide an accessible
illustration of the validity of the Lindblad equation to non-specialists.
The only prerequisite to follow the arguments exposed here is a basic
knowledge of quantum mechanics, including a familiarity with the concepts
of the density operator and the Liouville-von Neumann equation, at
the level of Ref. \cite{key-10}, for example.

The paper is structured as follows: in the Sec. II we derive the Born-Markov
master equation by tracing out the degrees of freedom of system $B$,
starting from the Liouville-von Neumann equation; in Sec. III we derive
the Lindblad equation; and in Sec. IV we present the conclusion.

\section{The Born-Markov master equation}

Let us consider a physical situation where a \emph{principal} system
{\emph{$S$}, whose dynamics is the object of interest, is coupled
with another quantum system \emph{$B$,} called \emph{bath}. Here,
$\mathcal{H_{S}}$ and $\mathcal{H_{B}}$ are, respectively, the Hilbert
spaces of principal system \emph{$S$} and bath\emph{$B$}; the global
Hilbert space \emph{$S+B$} will be represented by the tensor-product
space $\mathcal{H_{S}\otimes\mathcal{H_{B}}}$. The total Hamiltonian
is}

\begin{equation}
\hat{H}\left(t\right)=\hat{H}_{S}\otimes\hat{1}_{B}+\hat{1}_{S}\otimes\hat{H}_{B}+\alpha\hat{H}_{SB},
\end{equation}
 where $\hat{H}_{S}$ describes the principal system $S$, $\hat{H}_{B}$
describes the bath $B$, $\hat{H}_{SB}$ is the Hamiltonian for the
system-bath interaction and $\hat{1}_{B}$ and $\hat{1}_{S}$ are
the corresponding identities in the Hilbert spaces. Here, we will
considerate $\hat{H}_{S}$ and $\hat{H}_{B}$ both time-independent.
For the sake of simplicity, let us ignore the symbol $\otimes$ and
write 
\begin{equation}
\hat{H}\left(t\right)=\hat{H}_{S}+\hat{H}_{B}+\alpha\hat{H}_{SB}.\label{hamil}
\end{equation}
 Here, $\alpha$ is a real constant that provides the intensity of
interaction between the principal system and the bath. Writing $\hat{\rho}_{SB}$
for the global density operator ($S+B$), the Liouville-von Neumann
equation will be:

\begin{equation}
\frac{d}{dt}\hat{\rho}_{SB}=-\frac{i}{\hbar}\left[\hat{H}_{S}+\hat{H}_{B}+\alpha\hat{H}_{SB},\hat{\rho}_{SB}\right].\label{liouvon1}
\end{equation}
 It is convenient to write Eq. (\ref{liouvon1}) in the interaction
picture of $\hat{H}_{S}+\hat{H}_{B}$. With the definitions of the
new density operator and Hamiltonian:

\begin{equation}
\hat{H}\left(t\right)=e^{\frac{i}{\hbar}\left(\hat{H}_{S}+\hat{H}_{B}\right)t}\hat{H}_{SB}e^{-\frac{i}{\hbar}\left(\hat{H}_{S}+\hat{H}_{B}\right)t}\label{int}
\end{equation}
 and

\begin{equation}
\hat{\rho}\left(t\right)=e^{\frac{i}{\hbar}\left(\hat{H}_{S}+\hat{H}_{B}\right)t}\hat{\rho}_{SB}\left(t\right)e^{-\frac{i}{\hbar}\left(\hat{H}_{S}+\hat{H}_{B}\right)t},\label{rotrans}
\end{equation}
 the new equation for$\hat{\rho}\left(t\right)$ will be

\begin{equation}
\frac{d}{dt}\hat{\rho}\left(t\right)=-\frac{i}{\hbar}\alpha\left[\hat{H}\left(t\right),\hat{\rho}\left(t\right)\right].\label{liouvon2}
\end{equation}
 \textcolor{red}{{} }Here and in the following, we will use the time
argument explicited ($\left(t\right)$) to indicate the interaction-picture
transformation.

We want to find the evolution for $\hat{\rho}_{S}\left(t\right)=tr_{B}\left\{ \hat{\rho}_{SB}\left(t\right)\right\} $
where, according Eq. (\ref{rotrans}),

\begin{equation}
\hat{\rho}_{SB}\left(t\right)=e^{-\frac{i}{\hbar}\left(\hat{H}_{S}+\hat{H}_{B}\right)t}\hat{\rho}e^{\frac{i}{\hbar}\left(\hat{H}_{S}+\hat{H}_{B}\right)t}.
\end{equation}

Equation (\ref{liouvon2}) is the starting point of our iterative
approach. Its time derivative yields

\begin{equation}
\hat{\rho}\left(t\right)=\hat{\rho}\left(0\right)-\frac{i}{\hbar}\alpha\int_{0}^{t}\left[\hat{H}\left(t'\right),\hat{\rho}\left(t'\right)\right]dt'.\label{rint1}
\end{equation}
 Replacing Eq. (\ref{rint1}) into Eq. (\ref{liouvon2}), we have
\begin{equation}
\frac{d}{dt}\hat{\rho}\left(t\right)=-\frac{i}{\hbar}\alpha\left[\hat{H}\left(t\right),\hat{\rho}\left(0\right)\right]-\frac{1}{\hbar^{2}}\alpha^{2}\left[\hat{H}\left(t\right),\int_{0}^{t}\left[\hat{H}\left(t'\right),\hat{\rho}\left(t'\right)\right]dt'\right].\label{liouvon3}
\end{equation}
 For the\emph{ Born} \emph{approximation}, Eq. (\ref{liouvon3}) is
enough. Then, we take the partial trace of the bath degrees of freedom,

\begin{equation}
\frac{d}{dt}\hat{\rho}_{S}\left(t\right)=-\frac{i}{\hbar}\alpha tr_{B}\left\{ \left[\hat{H}\left(t\right),\hat{\rho}\left(0\right)\right]\right\} -\frac{1}{\hbar^{2}}\alpha^{2}tr_{B}\left\{ \left[\hat{H}\left(t\right),\int_{0}^{t}\left[\hat{H}\left(t'\right),\hat{\rho}\left(t'\right)\right]dt'\right]\right\} .\label{liouvon4}
\end{equation}

By the definition in Eq. (\ref{int}), $\hat{H}\left(t\right)$ depends
on $\hat{H}_{SB}$, and $\hat{H}_{SB}$ can always be defined in a
manner in which the first term of the right hand side of Eq. (\ref{liouvon4})
is zero. Hence, we obtain

\begin{equation}
\frac{d}{dt}\hat{\rho}_{S}\left(t\right)=-\frac{1}{\hbar^{2}}\alpha^{2}tr_{B}\left\{ \left[\hat{H}\left(t\right),\int_{0}^{t}\left[\hat{H}\left(t'\right),\hat{\rho}\left(t'\right)\right]dt'\right]\right\} .\label{liouvon5}
\end{equation}
 Integrating Eq. (\ref{liouvon5}) from $t$ to $t'$ yields

\begin{eqnarray*}
\hat{\rho}_{S}\left(t\right)-\hat{\rho}_{S}\left(t'\right) & = & -\frac{1}{\hbar^{2}}\alpha^{2}\int_{t}^{t'}dt'tr_{B}\left\{ \left[\hat{H}\left(t'\right),\int_{0}^{t'}\left[\hat{H}\left(t''\right),\hat{\rho}\left(t''\right)\right]dt''\right]\right\} ,
\end{eqnarray*}
 which shows that the difference between $\hat{\rho}_{S}\left(t\right)$
and $\hat{\rho}_{S}\left(t'\right)$ is of the second order of magnitude
in $\alpha$ and, therefore, we can write $\hat{\rho}_{S}\left(t\right)$
in the integrand of Eq. (\ref{liouvon5}), obtaining a time-local
equation for the density operator, without violating the Born approximation:
\begin{equation}
\frac{d}{dt}\hat{\rho}_{S}\left(t\right)=-\frac{1}{\hbar^{2}}\alpha^{2}tr_{B}\left\{ \left[\hat{H}\left(t\right),\int_{0}^{t}\left[\hat{H}\left(t'\right),\hat{\rho}\left(t\right)\right]dt'\right]\right\} .\label{liouvon6}
\end{equation}

The $\alpha$ constant was introduced in Eq. (\ref{hamil}) only for
clarifying the order of magnitude of each term in the iteration and,
now, it can be supressed, that is, let us take $\alpha=1$ (full interaction).
Thus, let us write: 
\begin{equation}
\frac{d}{dt}\hat{\rho}_{S}\left(t\right)=-\frac{1}{\hbar^{2}}tr_{B}\left\{ \left[\hat{H}\left(t\right),\int_{0}^{t}\left[\hat{H}\left(t'\right),\hat{\rho}\left(t\right)\right]dt'\right]\right\} .\label{liouvon7}
\end{equation}
 For this approximation, we can write $\hat{\rho}\left(t\right)=\hat{\rho}_{S}\left(t\right)\otimes\hat{\rho}_{B}$
inside the integral and obtain the equation that will be used in the
next calculations ({again, for the sake of simplicity, let us ignore
the symbol $\otimes$}): 
\begin{equation}
\frac{d}{dt}\hat{\rho}_{S}\left(t\right)=-\frac{1}{\hbar^{2}}\int_{0}^{\infty}dt'tr_{B}\left\{ \left[\hat{H}\left(t\right),\left[\hat{H}\left(t'\right),\hat{\rho}_{S}\left(t\right)\hat{\rho}_{B}\right]\right]\right\} ,\label{bornmarkovfinal}
\end{equation}
 where we assume that the integration can be extended to infinity
without changing its result. Equation (\ref{bornmarkovfinal}) is
the \emph{Born-Markov master equation}\cite{key-2}.

\section{Lindblad equation}

\subsection{The master equation commutator}

Let us consider that system-bath interaction is of the following form,

\begin{eqnarray}
\hat{H}_{SB} & = & \hbar\left(\hat{S}\hat{B}^{\dagger}+\hat{S}^{\dagger}\hat{B}\right),\label{defham}
\end{eqnarray}
 where $\hat{S}$ is a general operator that acts only on the principal
system $S$, and $\hat{B}$ is an operator that acts only on the bath
$B$. Now, we consider that $\hat{S}$ commutes with $\hat{H}_{S}$,
i. e., 
\begin{eqnarray*}
\left[\hat{S},\hat{H}_{S}\right] & = & 0,
\end{eqnarray*}
 resulting in 
\begin{eqnarray}
\hat{S}\left(t\right) & = & \hat{S}\label{propS}
\end{eqnarray}
 ($\hat{S}$ is not affected by the interaction-picture transformation).
Let us consider the bath hamiltonian defined by a bath of bosons,
\begin{equation}
\hat{H}_{B}=\hbar\underset{k}{\sum}\omega_{k}\hat{a}_{k}^{\dagger}\hat{a}_{k}
\end{equation}
 where $\hat{a}_{k}$ e $\hat{a}_{k}^{\dagger}$ are the annihilation
and creation bath operators, the $\omega_{k}$ are the characteristic
frequencies of each mode, and the $\hat{B}$ operator on Eq. (\ref{defham})
defined by

\begin{eqnarray}
\hat{B} & = & \sum_{k}g_{k}^{*}\hat{a}_{k},\label{B}
\end{eqnarray}
 where $g_{k}$ are complex coefficients representing coupling constantes.
Then, in the interaction picture, 
\begin{eqnarray}
\hat{B}\left(t\right) & = & e^{\frac{i}{\hbar}\hat{H}_{B}t}\hat{B}e^{-\frac{i}{\hbar}\hat{H}_{B}t}.\label{Bt}
\end{eqnarray}
 Expanding each exponential and using the commutator relations, Eq.
(\ref{Bt}) will result in

\begin{equation}
\hat{B}\left(t\right)=\sum_{k}g_{k}^{*}\hat{a}_{k}\mathrm{e}^{-i\omega_{k}t}.\label{Bt2}
\end{equation}
 The interaction (\ref{defham}) with the definition (\ref{B}) resembles
the \emph{Jaynes-Cummings} one, who represents a single two-level
atom interacting with a single mode of the radiation field \cite{key-2,key-11}.

With this, the commutator in Eq. (\ref{bornmarkovfinal}), $\left[\hat{H}\left(t\right),\left[\hat{H}\left(t'\right),\hat{\rho}_{S}\left(t\right)\hat{\rho}_{B}\right]\right]$,
will be evaluated. Firstly, 
\begin{eqnarray}
\left[\hat{H}\left(t\right),\left[\hat{H}\left(t^{\prime}\right),\hat{\rho}_{B}\hat{\rho}_{S}\left(t\right)\right]\right] & = & \hbar\left[\hat{S}\hat{B}^{\dagger}\left(t\right)+\hat{S}^{\dagger}\hat{B}\left(t\right),\left[\hat{H}\left(t^{\prime}\right),\hat{\rho}_{B}\hat{\rho}_{S}\left(t\right)\right]\right]\nonumber \\
 & = & \hbar\left[\hat{S}\hat{B}^{\dagger}\left(t\right),\left[\hat{H}\left(t^{\prime}\right),\hat{\rho}_{B}\hat{\rho}_{S}\left(t\right)\right]\right]\nonumber \\
 & + & \hbar\left[\hat{S}^{\dagger}\hat{B}\left(t\right),\left[\hat{H}\left(t^{\prime}\right),\hat{\rho}_{B}\hat{\rho}_{S}\left(t\right)\right]\right].\label{comt1}
\end{eqnarray}
 The gradual expansion of each term in Eq. (\ref{comt1}) will result
in

\begin{eqnarray}
\left[\hat{S}\hat{B}^{\dagger}\left(t\right),\left[\hat{H}\left(t^{\prime}\right),\hat{\rho}_{B}\hat{\rho}_{S}\left(t\right)\right]\right] & = & \hbar\left[\hat{S}\hat{B}^{\dagger}\left(t\right),\left[\hat{S}\hat{B}^{\dagger}\left(t'\right)+\hat{S}^{\dagger}\hat{B}\left(t'\right),\hat{\rho}_{B}\hat{\rho}_{S}\left(t\right)\right]\right]\nonumber \\
 & = & \hbar\hat{S}\hat{B}^{\dagger}\left(t\right)\left[\hat{S}\hat{B}^{\dagger}\left(t'\right)+\hat{S}^{\dagger}\hat{B}\left(t'\right)\right]\hat{\rho}_{B}\hat{\rho}_{S}\left(t\right)\nonumber \\
 & - & \hbar\hat{S}\hat{B}^{\dagger}\left(t\right)\hat{\rho}_{B}\hat{\rho}_{S}\left(t\right)\left[\hat{S}\hat{B}^{\dagger}\left(t'\right)+\hat{S}^{\dagger}\hat{B}\left(t'\right)\right]\nonumber \\
 & - & \hbar\left[\hat{S}\hat{B}^{\dagger}\left(t'\right)+\hat{S}^{\dagger}\hat{B}\left(t'\right)\right]\hat{\rho}_{B}\hat{\rho}_{S}\left(t\right)\hat{S}\hat{B}^{\dagger}\left(t\right)\nonumber \\
 & + & \hbar\hat{\rho}_{B}\hat{\rho}_{S}\left(t\right)\left[\hat{S}\hat{B}^{\dagger}\left(t'\right)+\hat{S}^{\dagger}\hat{B}\left(t'\right)\right]\hat{S}\hat{B}^{\dagger}\left(t\right)\label{comt2}
\end{eqnarray}
 and

\begin{eqnarray}
\left[\hat{S}^{\dagger}\hat{B}\left(t\right),\left[\hat{H}\left(t^{\prime}\right),\hat{\rho}_{B}\hat{\rho}_{S}\left(t\right)\right]\right] & = & \hbar\left[\hat{S}^{\dagger}\hat{B}\left(t\right),\left[\hat{S}\hat{B}^{\dagger}\left(t'\right)+\hat{S}^{\dagger}\hat{B}\left(t'\right),\hat{\rho}_{B}\hat{\rho}_{S}\left(t\right)\right]\right]\nonumber \\
 & = & \hbar\hat{S}^{\dagger}\hat{B}\left(t\right)\left[\hat{S}\hat{B}^{\dagger}\left(t'\right)+\hat{S}^{\dagger}\hat{B}\left(t'\right)\right]\hat{\rho}_{B}\hat{\rho}_{S}\left(t\right)\nonumber \\
 & - & \hbar\hat{S}^{\dagger}\hat{B}\left(t\right)\hat{\rho}_{B}\hat{\rho}_{S}\left(t\right)\left[\hat{S}\hat{B}^{\dagger}\left(t'\right)+\hat{S}^{\dagger}\hat{B}\left(t'\right)\right]\nonumber \\
 & - & \hbar\left[\hat{S}\hat{B}^{\dagger}\left(t'\right)+\hat{S}^{\dagger}\hat{B}\left(t'\right)\right]\hat{\rho}_{B}\hat{\rho}_{S}\left(t\right)\hat{S}^{\dagger}\hat{B}\left(t\right)\nonumber \\
 & + & \hbar\hat{\rho}_{B}\hat{\rho}_{S}\left(t\right)\left[\hat{S}\hat{B}^{\dagger}\left(t'\right)+\hat{S}^{\dagger}\hat{B}\left(t'\right)\right]\hat{S}^{\dagger}\hat{B}\left(t\right),\label{comt3}
\end{eqnarray}
 or, expanding Eqs. (\ref{comt2}) and (\ref{comt3}) and grouping
the similar terms in $S$ and $B$, we have:

\begin{eqnarray}
\left[\hat{S}\hat{B}^{\dagger}\left(t\right),\left[\hat{H}\left(t^{\prime}\right),\hat{\rho}_{B}\hat{\rho}_{S}\left(t\right)\right]\right] & = & \hbar\hat{S}\hat{S}\hat{\rho}_{S}\left(t\right)\hat{B}^{\dagger}\left(t\right)\hat{B}^{\dagger}\left(t'\right)\hat{\rho}_{B}+\hbar\hat{S}\hat{S}^{\dagger}\hat{\rho}_{S}\left(t\right)\hat{B}^{\dagger}\left(t\right)\hat{B}\left(t'\right)\hat{\rho}_{B}\nonumber \\
 & - & \hbar\hat{S}\hat{\rho}_{S}\left(t\right)\hat{S}\hat{B}^{\dagger}\left(t\right)\hat{\rho}_{B}\hat{B}^{\dagger}\left(t'\right)-\hbar\hat{S}\hat{\rho}_{S}\left(t\right)\hat{S}^{\dagger}\hat{B}^{\dagger}\left(t\right)\hat{\rho}_{B}\hat{B}\left(t'\right)\nonumber \\
 & - & \hbar\hat{S}\hat{\rho}_{S}\left(t\right)\hat{S}\hat{B}^{\dagger}\left(t'\right)\hat{\rho}_{B}\hat{B}^{\dagger}\left(t\right)-\hbar\hat{S}^{\dagger}\hat{\rho}_{S}\left(t\right)\hat{S}\hat{B}\left(t'\right)\hat{\rho}_{B}\hat{B}^{\dagger}\left(t\right)\nonumber \\
 & + & \hbar\hat{\rho}_{S}\left(t\right)\hat{S}\hat{S}\hat{\rho}_{B}\hat{B}^{\dagger}\left(t'\right)\hat{B}^{\dagger}\left(t\right)+\hbar\hat{\rho}_{S}\left(t\right)\hat{S}^{\dagger}\hat{S}\hat{\rho}_{B}\hat{B}\left(t'\right)\hat{B}^{\dagger}\left(t\right)\label{comt5}
\end{eqnarray}
 and

\begin{eqnarray}
\left[\hat{S}^{\dagger}\hat{B}\left(t\right),\left[\hat{H}\left(t^{\prime}\right),\hat{\rho}_{B}\hat{\rho}_{S}\left(t\right)\right]\right] & = & \hbar\hat{S}^{\dagger}\hat{S}\hat{\rho}_{S}\left(t\right)\hat{B}\left(t\right)\hat{B}^{\dagger}\left(t'\right)\hat{\rho}_{B}+\hbar\hat{S}^{\dagger}\hat{S}^{\dagger}\hat{\rho}_{S}\left(t\right)\hat{B}\left(t\right)\hat{B}\left(t'\right)\hat{\rho}_{B}\nonumber \\
 & - & \hbar\hat{S}^{\dagger}\hat{\rho}_{S}\left(t\right)\hat{S}\hat{B}\left(t\right)\hat{\rho}_{B}\hat{B}^{\dagger}\left(t'\right)-\hbar\hat{S}^{\dagger}\hat{\rho}_{S}\left(t\right)\hat{S}^{\dagger}\hat{B}\left(t\right)\hat{\rho}_{B}\hat{B}\left(t'\right)\nonumber \\
 & - & \hbar^{2}\hat{S}\hat{\rho}_{S}\left(t\right)\hat{S}^{\dagger}\hat{B}^{\dagger}\left(t'\right)\hat{\rho}_{B}\hat{B}\left(t\right)-\hbar\hat{S}^{\dagger}\hat{\rho}_{S}\left(t\right)\hat{S}^{\dagger}\hat{B}\left(t'\right)\hat{\rho}_{B}\hat{B}\left(t\right)\nonumber \\
 & + & \hbar\hat{\rho}_{S}\left(t\right)\hat{S}\hat{S}^{\dagger}\hat{\rho}_{B}\hat{B}^{\dagger}\left(t'\right)\hat{B}\left(t\right)+\hbar\hat{\rho}_{S}\left(t\right)\hat{S}^{\dagger}\hat{S}^{\dagger}\hat{\rho}_{B}\hat{B}\left(t'\right)\hat{B}\left(t\right).\label{comt6}
\end{eqnarray}

\subsection{The partial trace}

Now we are in a position to trace out the bath degrees of freedom
in Eqs. (\ref{comt5}) and (\ref{comt6}). As we can verify with Eq.
(\ref{Bt2}), 
\begin{eqnarray*}
tr_{B}\left\{ \hat{B}\left(t\right)\hat{B}\left(t^{\prime}\right)\hat{\rho}_{B}\right\}  & = & tr_{B}\left\{ \hat{B}^{\dagger}\left(t\right)\hat{B}^{\dagger}\left(t^{\prime}\right)\hat{\rho}_{B}\right\} =0,\:\forall t,t'.
\end{eqnarray*}

With this, then,

\begin{eqnarray}
tr_{B}\left\{ \left[\hat{S}\hat{B}^{\dagger}\left(t\right),\left[\hat{H}\left(t^{\prime}\right),\hat{\rho}_{B}\hat{\rho}_{S}\left(t\right)\right]\right]\right\}  & = & \hbar\hat{S}\hat{S}^{\dagger}\hat{\rho}_{S}\left(t\right)tr_{B}\left\{ \hat{B}^{\dagger}\left(t\right)\hat{B}\left(t'\right)\hat{\rho}_{B}\right\} \nonumber \\
 & - & \hbar\hat{S}\hat{\rho}_{S}\left(t\right)\hat{S}^{\dagger}tr_{B}\left\{ \hat{B}^{\dagger}\left(t\right)\hat{\rho}_{B}\hat{B}\left(t'\right)\right\} \nonumber \\
 & - & \hbar\hat{S}^{\dagger}\hat{\rho}_{S}\left(t\right)\hat{S}tr_{B}\left\{ \hat{B}\left(t'\right)\hat{\rho}_{B}\hat{B}^{\dagger}\left(t\right)\right\} \nonumber \\
 & + & \hbar\hat{\rho}_{S}\left(t\right)\hat{S}^{\dagger}\hat{S}tr_{B}\left\{ \hat{\rho}_{B}\hat{B}\left(t'\right)\hat{B}^{\dagger}\left(t\right)\right\} \label{comt7}
\end{eqnarray}
 and 
\begin{eqnarray}
tr_{B}\left\{ \left[\hat{S}^{\dagger}\hat{B}\left(t\right),\left[\hat{H}\left(t^{\prime}\right),\hat{\rho}_{B}\hat{\rho}_{S}\left(t\right)\right]\right]\right\}  & = & \hbar\hat{S}^{\dagger}\hat{S}\hat{\rho}_{S}\left(t\right)tr_{B}\left\{ \hat{B}\left(t\right)\hat{B}^{\dagger}\left(t'\right)\hat{\rho}_{B}\right\} \nonumber \\
 & - & \hbar\hat{S}^{\dagger}\hat{\rho}_{S}\left(t\right)\hat{S}tr_{B}\left\{ \hat{B}\left(t\right)\hat{\rho}_{B}\hat{B}^{\dagger}\left(t'\right)\right\} \nonumber \\
 & - & \hbar\hat{S}\hat{\rho}_{S}\left(t\right)\hat{S}^{\dagger}tr_{B}\left\{ \hat{B}^{\dagger}\left(t'\right)\hat{\rho}_{B}\hat{B}\left(t\right)\right\} \nonumber \\
 & + & \hbar\hat{\rho}_{S}\left(t\right)\hat{S}\hat{S}^{\dagger}tr_{B}\left\{ \hat{\rho}_{B}\hat{B}^{\dagger}\left(t'\right)\hat{B}\left(t\right)\right\} ,\label{comt8}
\end{eqnarray}
 where, if we use the ciclic properties of the trace,

\begin{eqnarray}
tr_{B}\left\{ \left[\hat{S}\hat{B}^{\dagger}\left(t\right),\left[\hat{H}\left(t^{\prime}\right),\hat{\rho}_{B}\hat{\rho}_{S}\left(t\right)\right]\right]\right\}  & = & \hbar\left[\hat{S}\hat{S}^{\dagger}\hat{\rho}_{S}\left(t\right)-\hat{S}^{\dagger}\hat{\rho}_{S}\left(t\right)\hat{S}\right]tr_{B}\left\{ \hat{B}^{\dagger}\left(t\right)\hat{B}\left(t'\right)\hat{\rho}_{B}\right\} \nonumber \\
 & + & \hbar\left[\hat{\rho}_{S}\left(t\right)\hat{S}^{\dagger}\hat{S}-\hat{S}\hat{\rho}_{S}\left(t\right)\hat{S}^{\dagger}\right]tr_{B}\left\{ \hat{B}\left(t'\right)\hat{B}^{\dagger}\left(t\right)\hat{\rho}_{B}\right\} \label{comt9}
\end{eqnarray}
 and

\begin{eqnarray}
tr_{B}\left\{ \left[\hat{S}^{\dagger}\hat{B}\left(t\right),\left[\hat{H}\left(t^{\prime}\right),\hat{\rho}_{B}\hat{\rho}_{S}\left(t\right)\right]\right]\right\}  & = & \hbar\left[\hat{S}^{\dagger}\hat{S}\hat{\rho}_{S}\left(t\right)-\hat{S}\hat{\rho}_{S}\left(t\right)\hat{S}^{\dagger}\right]tr_{B}\left\{ \hat{B}\left(t\right)\hat{B}^{\dagger}\left(t'\right)\hat{\rho}_{B}\right\} \nonumber \\
 & + & \hbar\left[\hat{\rho}_{S}\left(t\right)\hat{S}\hat{S}^{\dagger}-\hat{S}^{\dagger}\hat{\rho}_{S}\left(t\right)\hat{S}\right]tr_{B}\left\{ \hat{B}^{\dagger}\left(t'\right)\hat{B}\left(t\right)\hat{\rho}_{B}\right\} .\label{comt10}
\end{eqnarray}

The terms represented by Eqs. (\ref{comt9}) and (\ref{comt10}) allow
us to return to the Eq. (\ref{comt1}):

\begin{eqnarray}
tr_{B}\left\{ \left[\hat{H}\left(t\right),\left[\hat{H}\left(t^{\prime}\right),\hat{\rho}_{B}\hat{\rho}_{S}\left(t\right)\right]\right]\right\}  & = & \hbar^{2}\left[\hat{S}\hat{S}^{\dagger}\hat{\rho}_{S}\left(t\right)-\hat{S}^{\dagger}\hat{\rho}_{S}\left(t\right)\hat{S}\right]tr_{B}\left\{ \hat{B}^{\dagger}\left(t\right)\hat{B}\left(t'\right)\hat{\rho}_{B}\right\} \nonumber \\
 & + & \hbar^{2}\left[\hat{\rho}_{S}\left(t\right)\hat{S}^{\dagger}\hat{S}-\hat{S}\hat{\rho}_{S}\left(t\right)\hat{S}^{\dagger}\right]tr_{B}\left\{ \hat{B}\left(t'\right)\hat{B}^{\dagger}\left(t\right)\hat{\rho}_{B}\right\} \nonumber \\
 & + & \hbar^{2}\left[\hat{S}^{\dagger}\hat{S}\hat{\rho}_{S}\left(t\right)-\hat{S}\hat{\rho}_{S}\left(t\right)\hat{S}^{\dagger}\right]tr_{B}\left\{ \hat{B}\left(t\right)\hat{B}^{\dagger}\left(t'\right)\hat{\rho}_{B}\right\} \nonumber \\
 & + & \hbar^{2}\left[\hat{\rho}_{S}\left(t\right)\hat{S}\hat{S}^{\dagger}-\hat{S}^{\dagger}\hat{\rho}_{S}\left(t\right)\hat{S}\right]tr_{B}\left\{ \hat{B}^{\dagger}\left(t'\right)\hat{B}\left(t\right)\hat{\rho}_{B}\right\} .\label{comt11}
\end{eqnarray}

\subsection{The expansion of the integrand of the master equation}

With the results of the preceding paragraphs, the integrand in Eq.
(\ref{bornmarkovfinal}) becomes:

\begin{eqnarray}
tr_{B}\left\{ \left[\hat{H}\left(t\right),\left[\hat{H}\left(t^{\prime}\right),\hat{\rho}_{B}\hat{\rho}_{S}\left(t\right)\right]\right]\right\}  & = & \hbar^{2}\left[\hat{S}\hat{S}^{\dagger}\hat{\rho}_{S}\left(t\right)-\hat{S}^{\dagger}\hat{\rho}_{S}\left(t\right)\hat{S}\right]tr_{B}\left\{ \hat{B}^{\dagger}\left(t\right)\hat{B}\left(t'\right)\hat{\rho}_{B}\right\} \nonumber \\
 & + & \hbar^{2}\left[\hat{\rho}_{S}\left(t\right)\hat{S}^{\dagger}\hat{S}-\hat{S}\hat{\rho}_{S}\left(t\right)\hat{S}^{\dagger}\right]tr_{B}\left\{ \hat{B}\left(t'\right)\hat{B}^{\dagger}\left(t\right)\hat{\rho}_{B}\right\} \nonumber \\
 & + & \hbar^{2}\left[\hat{S}^{\dagger}\hat{S}\hat{\rho}_{S}\left(t\right)-\hat{S}\hat{\rho}_{S}\left(t\right)\hat{S}^{\dagger}\right]tr_{B}\left\{ \hat{B}\left(t\right)\hat{B}^{\dagger}\left(t'\right)\hat{\rho}_{B}\right\} \nonumber \\
 & + & \hbar^{2}\left[\hat{\rho}_{S}\left(t\right)\hat{S}\hat{S}^{\dagger}-\hat{S}^{\dagger}\hat{\rho}_{S}\left(t\right)\hat{S}\right]tr_{B}\left\{ \hat{B}^{\dagger}\left(t'\right)\hat{B}\left(t\right)\hat{\rho}_{B}\right\} .\label{comt12}
\end{eqnarray}
 For convenience, let us define the functions

\begin{eqnarray}
F\left(t\right) & = & \int_{0}^{t}dt^{\prime}tr_{B}\left\{ \hat{B}\left(t\right)\hat{B}^{\dagger}\left(t^{\prime}\right)\hat{\rho}_{B}\right\} ,\nonumber \\
G\left(t\right) & = & \int_{0}^{t}dt^{\prime}tr_{B}\left\{ \hat{B}^{\dagger}\left(t^{\prime}\right)\hat{B}\left(t\right)\rho_{B}\right\} .\label{FGgeral}
\end{eqnarray}
 Then, 
\begin{eqnarray*}
F^{*}\left(t\right) & = & \int_{0}^{t}dt^{\prime}tr_{B}\left\{ \hat{B}\left(t^{\prime}\right)\hat{B}^{\dagger}\left(t\right)\hat{\rho}_{B}\right\} ,\\
G^{*}\left(t\right) & = & \int_{0}^{t}dt^{\prime}tr_{B}\left\{ \hat{B}^{\dagger}\left(t\right)\hat{B}\left(t^{\prime}\right)\hat{\rho}_{B}\right\} .
\end{eqnarray*}
 Replacing Eq. (\ref{comt12}) in Eq. (\ref{bornmarkovfinal}) yields:

\begin{eqnarray}
\frac{d}{dt}\hat{\rho}_{S}\left(t\right) & = & -\left[\hat{S}\hat{S}^{\dagger}\hat{\rho}_{S}\left(t\right)-\hat{S}^{\dagger}\hat{\rho}_{S}\left(t\right)\hat{S}\right]G^{*}\left(t\right)-\left[\hat{\rho}_{S}\left(t\right)\hat{S}^{\dagger}\hat{S}-\hat{S}\hat{\rho}_{S}\left(t\right)\hat{S}^{\dagger}\right]F^{*}\left(t\right)\nonumber \\
 & - & \left[\hat{S}^{\dagger}\hat{S}\hat{\rho}_{S}\left(t\right)-\hat{S}\hat{\rho}_{S}\left(t\right)\hat{S}^{\dagger}\right]F\left(t\right)-\left[\hat{\rho}_{S}\left(t\right)\hat{S}\hat{S}^{\dagger}-\hat{S}^{\dagger}\hat{\rho}_{S}\left(t\right)\hat{S}\right]G\left(t\right).\label{eqtrac1}
\end{eqnarray}
 Actually, the usual Lindblad equation emerges when $G(t)=0$ and
$F(t)=F^{*}(t)$. In the following, we make some specifications about
the environment to discuss these approximations in detail.

\subsection{The bath specification}

Furthermore, for the initial state of the thermal bath, we consider
the vacuum state: 
\begin{equation}
\hat{\rho}_{B}=\left(\left|0\right\rangle \left|0\right\rangle ...\right)\otimes\left(\left\langle 0\right|\left\langle 0\right|...\right).\label{rBvacuo}
\end{equation}

The evaluation of the $F\left(t\right)$ and $G\left(t\right)$ functions
defined in Eq. (\ref{FGgeral}) are done considering the $\hat{B}\left(t\right)$
and $\hat{\rho}_{B}$ definitions in Eqs. (\ref{Bt2}) and (\ref{rBvacuo}).
By Eq. (\ref{Bt2}), $\hat{B}^{\dagger}\left(t\right)$ is

\begin{equation}
\hat{B}^{\dagger}\left(t\right)=\sum_{k}g_{k}\hat{a}_{k}^{\dagger}\mathrm{e}^{i\omega_{k}t}.\label{Btm}
\end{equation}
 Then, the partial trace in $F\left(t\right)$ and $G\left(t\right)$
can be evaluated:

\begin{eqnarray}
tr_{B}\left\{ \hat{B}\left(t\right)\hat{B}^{\dagger}\left(t^{\prime}\right)\hat{\rho}_{B}\right\}  & = & tr_{B}\left\{ \hat{B}\left(t\right)\hat{B}^{\dagger}\left(t^{\prime}\right)\left(\left|0\right\rangle \left|0\right\rangle ...\right)\otimes\left(\left\langle 0\right|\left\langle 0\right|...\right)\right\} \label{tracF1}
\end{eqnarray}
 and

\begin{eqnarray}
tr_{B}\left\{ \hat{B}^{\dagger}\left(t^{\prime}\right)\hat{B}\left(t\right)\hat{\rho}_{B}\right\}  & = & tr_{B}\left\{ \hat{B}^{\dagger}\left(t^{\prime}\right)\hat{B}\left(t\right)\left(\left|0\right\rangle \left|0\right\rangle ...\right)\otimes\left(\left\langle 0\right|\left\langle 0\right|...\right)\right\} .\label{tracG1}
\end{eqnarray}

If we use some bath state basis$\left\{ \left|b\right\rangle \right\} $,
Eqs. (\ref{tracF1}) and (\ref{tracG1}) become

\begin{eqnarray}
tr_{B}\left\{ \hat{B}\left(t\right)\hat{B}^{\dagger}\left(t^{\prime}\right)\hat{\rho}_{B}\right\}  & = & \underset{b}{\sum}\left\langle b\right|\hat{B}\left(t\right)\hat{B}^{\dagger}\left(t^{\prime}\right)\left(\left|0\right\rangle \left|0\right\rangle ...\right)\otimes\left(\left\langle 0\right|\left\langle 0\right|...\right)\left|b\right\rangle \nonumber \\
 & = & \underset{b}{\sum}\left(\left\langle 0\right|\left\langle 0\right|...\right)\left|b\right\rangle \left\langle b\right|\hat{B}\left(t\right)\hat{B}^{\dagger}\left(t^{\prime}\right)\left(\left|0\right\rangle \left|0\right\rangle ...\right)\nonumber \\
 & = & \left(\left\langle 0\right|\left\langle 0\right|...\right)\underset{b}{\sum}\left|b\right\rangle \left\langle b\right|\hat{B}\left(t\right)\hat{B}^{\dagger}\left(t^{\prime}\right)\left(\left|0\right\rangle \left|0\right\rangle ...\right)\nonumber \\
 & = & \left(\left\langle 0\right|\left\langle 0\right|...\right)\hat{B}\left(t\right)\hat{B}^{\dagger}\left(t^{\prime}\right)\left(\left|0\right\rangle \left|0\right\rangle ...\right)\label{tracF2}
\end{eqnarray}
 and

\begin{eqnarray}
tr_{B}\left\{ \hat{B}^{\dagger}\left(t^{\prime}\right)\hat{B}\left(t\right)\hat{\rho}_{B}\right\}  & = & \underset{b}{\sum}\left\langle b\right|\hat{B}^{\dagger}\left(t^{\prime}\right)\hat{B}\left(t\right)\left(\left|0\right\rangle \left|0\right\rangle ...\right)\otimes\left(\left\langle 0\right|\left\langle 0\right|...\right)\left|b\right\rangle \nonumber \\
 & = & \underset{b}{\sum}\left(\left\langle 0\right|\left\langle 0\right|...\right)\left|b\right\rangle \left\langle b\right|\hat{B}^{\dagger}\left(t^{\prime}\right)\hat{B}\left(t\right)\left(\left|0\right\rangle \left|0\right\rangle ...\right)\nonumber \\
 & = & \left(\left\langle 0\right|\left\langle 0\right|...\right)\underset{b}{\sum}\left|b\right\rangle \left\langle b\right|\hat{B}^{\dagger}\left(t^{\prime}\right)\hat{B}\left(t\right)\left(\left|0\right\rangle \left|0\right\rangle ...\right)\nonumber \\
 & = & \left(\left\langle 0\right|\left\langle 0\right|...\right)\hat{B}^{\dagger}\left(t^{\prime}\right)\hat{B}\left(t\right)\left(\left|0\right\rangle \left|0\right\rangle ...\right).\label{tracG2}
\end{eqnarray}
 Let us expand $\hat{B}^{\dagger}\left(t\right)$ and $\hat{B}\left(t\right)$
using Eqs. (\ref{Bt2}) and (\ref{Btm}): 
\begin{eqnarray}
tr_{B}\left\{ \hat{B}\left(t\right)\hat{B}^{\dagger}\left(t^{\prime}\right)\hat{\rho}_{B}\right\}  & = & \left(\left\langle 0\right|\left\langle 0\right|...\right)\sum_{k}g_{k}^{*}\hat{a}_{k}\mathrm{e}^{-i\omega_{k}t}\sum_{k'}g_{k'}\hat{a}_{k'}^{\dagger}\mathrm{e}^{i\omega_{k'}t'}\left(\left|0\right\rangle \left|0\right\rangle ...\right)\nonumber \\
 & = & \sum_{k,k'}g_{k}^{*}g_{k'}\mathrm{e}^{-i\left(\omega_{k}t-\omega_{k'}t'\right)}\left(\left\langle 0\right|\left\langle 0\right|...\right)\hat{a}_{k}\hat{a}_{k'}^{\dagger}\left(\left|0\right\rangle \left|0\right\rangle ...\right)\label{tracF3}
\end{eqnarray}
 and

\begin{eqnarray}
tr_{B}\left\{ \hat{B}^{\dagger}\left(t^{\prime}\right)\hat{B}\left(t\right)\rho_{B}\right\}  & = & \left(\left\langle 0\right|\left\langle 0\right|...\right)\sum_{k'}g_{k'}\hat{a}_{k'}^{\dagger}\mathrm{e}^{i\omega_{k'}t'}\sum_{k}g_{k}^{*}\hat{a}_{k}\mathrm{e}^{-i\omega_{k}t}\left(\left|0\right\rangle \left|0\right\rangle ...\right)\nonumber \\
 & = & \sum_{k,k'}g_{k}^{*}g_{k'}\mathrm{e}^{-i\left(\omega_{k}t-\omega_{k'}t'\right)}\left(\left\langle 0\right|\left\langle 0\right|...\right)\hat{a}_{k'}^{\dagger}\hat{a}_{k}\left(\left|0\right\rangle \left|0\right\rangle ...\right)=0\label{tracG3}
\end{eqnarray}
 Hence, we can rewrite Eq. (\ref{tracF3}) with the $\hat{a}_{k}^{\dagger}$
operators on the left of the $\hat{a}_{k}$ operators. We know that

\begin{equation}
\hat{a}_{k}\hat{a}_{k'}^{\dagger}=\delta_{k,k'}+\hat{a}_{k'}^{\dagger}\hat{a}_{k}.
\end{equation}
 Then

\begin{eqnarray}
tr_{B}\left\{ \hat{B}\left(t\right)\hat{B}^{\dagger}\left(t^{\prime}\right)\hat{\rho}_{B}\right\}  & = & \sum_{k,k'}g_{k}^{*}g_{k'}\mathrm{e}^{-i\left(\omega_{k}t-\omega_{k'}t'\right)}\delta_{k,k'}+\sum_{k,k'}g_{k}^{*}g_{k'}\mathrm{e}^{-i\left(\omega_{k}t-\omega_{k'}t'\right)}\left(\left\langle 0\right|\left\langle 0\right|...\right)\hat{a}_{k'}^{\dagger}\hat{a}_{k}\left(\left|0\right\rangle \left|0\right\rangle ...\right)\nonumber \\
 & = & \sum_{k}\left|g_{k}\right|^{2}\mathrm{e}^{-i\omega_{k}\left(t-t'\right)}.\label{tracF4}
\end{eqnarray}
 Therefore, from Eqs. (\ref{tracG3}) and (\ref{tracF4}), it follows
that 
\begin{eqnarray}
F\left(t\right) & = & \sum_{k}\left|g_{k}\right|^{2}\int_{0}^{t}dt^{\prime}\mathrm{e}^{-i\omega_{k}\left(t-t'\right)},\nonumber \\
G\left(t\right) & = & 0.\label{FGcalculados}
\end{eqnarray}

\subsection{Transition to the continuum}

In the expression of $F\left(t\right)$ in Eq. (\ref{FGcalculados}),
if we adopt the general definition of the density of states as

\begin{equation}
J\left(\omega\right)=\underset{l}{\sum}\left|g_{l}\right|^{2}\delta\left(\omega-\omega_{l}\right),
\end{equation}
 then the sum over $k$ can be replaced by an integral over a continuum
of frequencies: 
\begin{eqnarray*}
F\left(t\right) & = & \int_{0}^{\infty}d\omega J\left(\omega\right)\int_{0}^{t}dt^{\prime}\mathrm{e}^{-i\omega\left(t-t'\right)}.
\end{eqnarray*}
 Let us introduce the new variable 
\begin{eqnarray*}
\tau & = & t-t',\\
d\tau & = & -dt',
\end{eqnarray*}
 with

\begin{eqnarray*}
\int_{0}^{t}dt' & = & -\int_{t}^{0}d\tau=\int_{0}^{t}d\tau,
\end{eqnarray*}
 yielding 
\begin{eqnarray*}
F\left(t\right) & = & \int_{0}^{\infty}d\omega J\left(\omega\right)\int_{0}^{t}d\tau\mathrm{e}^{-i\omega\tau}.
\end{eqnarray*}

\subsection{The Markov approximation}

In the Markov approximation, the limit $t\rightarrow\infty$ is taken
in the time integral, as we have mentioned regarding Eq. (\ref{bornmarkovfinal}),
that is, we take 
\begin{eqnarray*}
 & \int_{0}^{\infty}d\tau\mathrm{e}^{-i\omega\tau}.
\end{eqnarray*}
 As the integrand \textcolor{red}{oscillates}, we will use the device:
\begin{eqnarray*}
\int_{0}^{\infty}d\tau\mathrm{e}^{-i\omega\tau} & = & \lim_{\eta\rightarrow0^{+}}\int_{0}^{\infty}d\tau\mathrm{e}^{-i\omega\tau-\eta\tau}\\
 & = & \lim_{\eta\rightarrow0^{+}}\frac{1}{\eta+i\omega}\\
 & = & \lim_{\eta\rightarrow0^{+}}\frac{\eta-i\omega}{\eta^{2}+\omega^{2}}\\
 & = & \lim_{\eta\rightarrow0^{+}}\frac{\eta}{\eta^{2}+\omega^{2}}-\lim_{\eta\rightarrow0^{+}}\frac{i\omega}{\eta^{2}+\omega^{2}}\\
 & = & \pi\delta\left(\omega\right)-i\mathrm{P}\frac{1}{\omega},
\end{eqnarray*}
 where $\mathrm{P}$ stands for the \emph{Cauchy principal part}.
Then, 
\begin{eqnarray*}
F & = & \pi\int_{0}^{\infty}d\omega J\left(\omega\right)\delta\left(\omega\right)-i\mathrm{P}\int_{0}^{\infty}d\omega\frac{J\left(\omega\right)}{\omega}.
\end{eqnarray*}

\subsection{The final form}

For a general density of states, $F$ yields

\begin{equation}
F=\frac{\gamma+i\varepsilon}{2},\label{Ffinal}
\end{equation}
 where

\begin{eqnarray}
\gamma & \equiv & 2\pi\int_{0}^{\infty}d\omega J\left(\omega\right)\delta\left(\omega\right),\\
\varepsilon & \equiv & -2\mathrm{P}\int_{0}^{\infty}d\omega\frac{J\left(\omega\right)}{\omega}.
\end{eqnarray}
 As we have verified that $G=0$, let us replace Eq. (\ref{Ffinal})
in Eq. (\ref{eqtrac1}): 
\begin{eqnarray*}
\frac{d}{dt}\hat{\rho}_{S}\left(t\right) & = & -\left[\hat{\rho}_{S}\left(t\right)\hat{S}^{\dagger}\hat{S}-\hat{S}\hat{\rho}_{S}\left(t\right)\hat{S}^{\dagger}\right]\frac{\gamma-i\varepsilon}{2}-\left[\hat{S}^{\dagger}\hat{S}\hat{\rho}_{S}\left(t\right)-\hat{S}\hat{\rho}_{S}\left(t\right)\hat{S}^{\dagger}\right]\frac{\gamma+i\varepsilon}{2}\\
 & = & -\frac{\gamma}{2}\left[\hat{\rho}_{S}\left(t\right)\hat{S}^{\dagger}\hat{S}-\hat{S}\hat{\rho}_{S}\left(t\right)\hat{S}^{\dagger}+\hat{S}^{\dagger}\hat{S}\hat{\rho}_{S}\left(t\right)-\hat{S}\hat{\rho}_{S}\left(t\right)\hat{S}^{\dagger}\right]\\
 & + & i\frac{\varepsilon}{2}\left[\hat{\rho}_{S}\left(t\right)\hat{S}^{\dagger}\hat{S}-\hat{S}\hat{\rho}_{S}\left(t\right)\hat{S}^{\dagger}-\hat{S}^{\dagger}\hat{S}\hat{\rho}_{S}\left(t\right)-\hat{S}\hat{\rho}_{S}\left(t\right)\hat{S}^{\dagger}\right].
\end{eqnarray*}
 If the density of states is chosen to yield $\varepsilon=0$ (a \emph{Lorentzian},
for example, where we can extend the lower limit of integration to
$-\infty$), the final result is:

\begin{equation}
\frac{d}{dt}\hat{\rho}_{S}\left(t\right)=\gamma\left[\hat{S}\hat{\rho}_{S}\left(t\right)\hat{S}^{\dagger}-\frac{1}{2}\left\{ \hat{S}^{\dagger}\hat{S},\hat{\rho}_{S}\left(t\right)\right\} \right].\label{eqlind1}
\end{equation}
 Let us, then, return to the original picture. Since 
\begin{equation}
\hat{\rho}_{S}\left(t\right)=e^{\frac{i}{\hbar}\hat{H}_{S}t}\hat{\rho}_{S}e^{-\frac{i}{\hbar}\hat{H}_{S}t},
\end{equation}
 then

\begin{eqnarray}
\frac{d}{dt}\hat{\rho}_{S}\left(t\right) & = & \frac{i}{\hbar}e^{\frac{i}{\hbar}\hat{H}_{S}t}\hat{H}_{S}\hat{\rho}_{S}e^{-\frac{i}{\hbar}\hat{H}_{S}t}+e^{\frac{i}{\hbar}\hat{H}_{S}t}\frac{d\hat{\rho}_{S}}{dt}e^{-\frac{i}{\hbar}\hat{H}_{S}t}-\frac{i}{\hbar}e^{\frac{i}{\hbar}\hat{H}_{S}t}\hat{\rho}_{S}\hat{H}_{S}e^{-\frac{i}{\hbar}\hat{H}_{S}t}\nonumber \\
 & = & e^{\frac{i}{\hbar}\hat{H}_{S}t}\frac{d\hat{\rho}_{S}}{dt}e^{-\frac{i}{\hbar}\hat{H}_{S}t}+\frac{i}{\hbar}e^{\frac{i}{\hbar}\hat{H}_{S}t}\left[\hat{H}_{S},\hat{\rho}_{S}\right]e^{-\frac{i}{\hbar}\hat{H}_{S}t}.\label{auxlind1}
\end{eqnarray}
 Performing the same operation on the right-hand side of Eq. (\ref{eqlind1})
gives

\begin{equation}
\left[\hat{S}\hat{\rho}_{S}\left(t\right)\hat{S}^{\dagger}-\frac{1}{2}\left\{ \hat{S}^{\dagger}\hat{S},\hat{\rho}_{S}\left(t\right)\right\} \right]=e^{\frac{i}{\hbar}\hat{H}_{S}t}\left[\hat{S}\hat{\rho}_{S}\hat{S}^{\dagger}-\frac{1}{2}\left\{ \hat{S}^{\dagger}\hat{S},\hat{\rho}_{S}\right\} \right]e^{-\frac{i}{\hbar}\hat{H}_{S}t}.\label{auxlind2}
\end{equation}
 Replacing Eqs. (\ref{auxlind1}) and (\ref{auxlind2}) in Eq. (\ref{eqlind1}),
we obtain:

\begin{equation}
\frac{d\hat{\rho}_{S}}{dt}=-\frac{i}{\hbar}\left[\hat{H}_{S},\hat{\rho}_{S}\right]+\gamma\left[\hat{S}\hat{\rho}_{S}\hat{S}^{\dagger}-\frac{1}{2}\left\{ \hat{S}^{\dagger}\hat{S},\hat{\rho}_{S}\right\} \right].\label{Lindblad}
\end{equation}

\section{Conclusion}

In summary, in this paper we consider an interaction that resembles
the \emph{Jaynes-Cummings interaction} \cite{key-2}, Eq. (\ref{defham}),
between a bath and a system $S$, assuming that the operator $\hat{S}$
commutes with the system Hamiltonian, $\hat{H}_{S}$, at zero temperature.
We substituted Eq. (\ref{defham}) in the Born-Markov master equation
(\ref{bornmarkovfinal}) and took the partial trace of the degrees
of freedom of $B$. The $T=0$ hypothesis is necessary to simplify
the calculations, making them more accessible to the students, simplifying
the treatment of Eqs. (\ref{tracF1}) and (\ref{tracG1}), and avoiding
complications such as the Lamb shift in Eq. (\ref{Ffinal}). The Markov
approximation in Sec. III-F was also vital to obtain the final result,
Eq. (\ref{Lindblad}). All these simplifications limit the validity
of our derivation to more general cases, but it provides a detailed
illustration of the physical meaning of each term appearing in the
Lindblad equation.

Equation (\ref{Lindblad}) is commonly presented with several $\hat{S}$
operators, usually denoted by $\hat{L}$, in a linear combination
of Lindbladian operators. The $\hat{L}$ operators are named \emph{Lindblad
operators} and, in the general case, the Lindblad equation takes the
form:

\begin{equation}
\frac{d\hat{\rho}_{S}}{dt}=-\frac{i}{\hbar}\left[\hat{H}_{S},\hat{\rho}_{S}\right]+\gamma\underset{j}{\sum}\left[\hat{L}_{j}\hat{\rho}_{S}\hat{L}_{j}^{\dagger}-\frac{1}{2}\left\{ \hat{L}_{j}^{\dagger}\hat{L}_{j},\hat{\rho}_{S}\right\} \right].\label{lindgeral}
\end{equation}
 If we consider only the first term on the right hand side of Eq.
(\ref{lindgeral}), we obtain the \emph{Liouville-von Neumann equation}.
This term is the \emph{Liouvillian} and describes the \emph{unitary
evolution} of the density operator. The second term on the right hand
side of the Eq. (\ref{Lindblad}) is the \emph{Lindbladian }and it
emerges when we take the partial trace - \emph{a non-unitary operation}
- of the degrees of freedom of system $B$. The Lindbladian describes
the \emph{non-unitary evolution} of the density operator. By the interaction
form adopted here, Eq. (\ref{defham}), the physical meaning of the
Lindblad operators can be understood: they represent the system $S$
contribution to the $S-B$ interaction - remembering once more that
the Lindblad equation was derived from the Liouville-von Neumann one
by tracing the bath degrees of freedom. This conclusion is also achieved
with the more general derivation \cite{key-1,key-2}. It is important
to emphasize that, due to our simplifying assumptions, the summation
appearing in Eq. (\ref{lindgeral}) was not obtained in our derivation
of Eq. (\ref{Lindblad}).

If the Lindblad operators $\hat{L}_{j}$ are Hermitian (observables),
the Lindblad equation can be used to treat the measurement process.
A simple application in this sense is the Hamiltonian $\hat{H}_{S}\propto\hat{\sigma}_{z}$
($\hat{\sigma}_{z}$ is the 2-level \emph{z}-Pauli mattrix) when we
want to measure one specific component of the spin ($\hat{L}\propto\hat{\sigma}_{\alpha},\:\alpha=x,y,z$,
without the summation) \cite{key-3,key-12}. If the Lindblads are
non-Hermitian, the equation can be used to treat dissipation, decoherence
or decays. For this, a simple example is the same Hamiltonian $\hat{H}_{S}\propto\hat{\sigma}_{z}$
with the Lindblad $\hat{L}\propto\hat{\sigma}_{-}$ ($\hat{\sigma}_{-}=\frac{\hat{\sigma}_{x}-i\hat{\sigma}_{y}}{2}$),
where $\gamma$ will be the spontaneous emission rate \cite{key-13}. 
\begin{acknowledgments}
C. A. Brasil acknowledges support from Coordenação de Aperfeiçoamento
de Pessoal de Nível Superior (CAPES) and Fundação de Amparo à Pesquisa
do Estado de São Paulo (FAPESP) project number 2011/19848-4, Brazil.

F. F. Fanchini acknowledges support from FAPESP and Conselho Nacional de
Desenvolvimento Científico e Tecnológico (CNPq) through the Instituto Nacional de Ciência e Tecnologia - Informação Quântica (INCT-IQ), Brazil.

R. d. J. Napolitano acknowledges support from CNPq, Brazil.\end{acknowledgments}

\end{document}